\documentclass{PoS}

\usepackage{bm}
\usepackage{amsmath,amssymb}

\newcommand{\im}{\text{Im }}

\title{Origin of resonances in chiral dynamics}

\ShortTitle{Origin of resonances in chiral dynamics}

\author{\speaker{Tetsuo Hyodo} \\
        Department of Physics, Tokyo Institute of Technology, 
	Meguro 152-8551, Japan \\
        E-mail: \email{hyodo@th.phys.titech.ac.jp}}

\author{Daisuke~Jido\\
        Yukawa Institute for Theoretical Physics,
Kyoto University, Kyoto 606-8502, Japan
}

	\author{Atsushi~Hosaka\\
        Research Center for Nuclear Physics (RCNP),
Ibaraki, Osaka 567-0047, Japan 
}

\abstract{The nature of baryon resonances is studied in the dynamical chiral 
coupled-channel approach for meson-baryon scattering. In general, origin of 
resonances in two-body scattering can be classified into two categories: 
dynamically generated states and genuine elementary particles. We demonstrate 
that the genuine contribution in the loop function can be excluded by adopting 
a natural renormalization scheme. The origin of resonances can be studied by 
looking at the effective interaction in the natural renormalization scheme, which
is deduced from the phenomenological amplitude fitted to experimental data. 
Applying this method to the baryon resonances, we find that the dominant 
component for the $\Lambda(1405)$ resonance is dynamical, while a genuine
contribution plays a substantial role for the structure of the $N(1535)$.}

\FullConference{6th International Workshop on Chiral Dynamics, CD09\\
		July 6-10, 2009\\
		Bern, Switzerland}

\begin{document}

\section{Introduction}\label{sec:intro}

One of the great successes of current algebra in 1960's is the celebrated 
Weinberg-Tomozawa theorem~\cite{Weinberg:1966kf,Tomozawa:1966jm}, which tells us 
that the chiral symmetry highly constrains the low energy $s$-wave scattering of
the Nambu-Goldstone boson with a target hadron. The establishment of power 
counting~\cite{Weinberg:1979kz} enables one to sort out the effective chiral
Lagrangian and amplitude, leading to the systematic computation of the higher 
order correction to current algebra, which results in the chiral perturbation 
theory~\cite{Gasser:1985gg}. In recent years, implementation of the chiral low 
energy interaction into the dynamical framework of hadron scattering was turned 
out to be a powerful tool to study resonance physics. This chiral coupled-channel
approach has been providing fairly successful description of baryon resonances in
the scattering of an octet pseudoscalar meson and an octet ground state 
baryon~\cite{Kaiser:1995eg,Oset:1998it,Oller:2000fj,Lutz:2001yb}. 

Once a good description of a resonance is obtained, we may next consider what 
structure it has. Excited baryons can consist of several components, such as 
three-quark state, meson-baryon molecule, and more complicated structures. 
Quantum theory tells us that the physical state must be realized as a 
superposition of all possible components, as far as they have the same quantum 
numbers. Hence, the clarification of the \textit{dominant} component among others
should help our intuitive understanding of the structure. 

In chiral dynamics, the excited baryons are described as resonances in the 
meson-baryon scattering amplitude. In this case, any components other than 
dynamical two-body state (meson-baryon molecule) are expressed by the 
Castillejo-Dalitz-Dyson (CDD) pole contribution~\cite{Castillejo:1956ed}.
Therefore, the estimation of the size of the CDD pole contribution in the 
amplitude will shed light on the origin of resonances in this approach. We would
like to report our recent study on the origin of baryon resonances in chiral 
dynamics, paying special attention to the renormalization 
procedure~\cite{Hyodo:2008xr}.

\section{Chiral dynamics for meson-baryon scattering}\label{sec:ChU}

We first write down the general form of the amplitude based on the N/D 
method~\cite{Oller:2000fj}, for $s$-wave single-channel meson-baryon scattering 
at total energy $\sqrt{s}$:
\begin{align}
    T(\sqrt{s};a)
    =& \frac{1}{V^{-1}(\sqrt{s})-G(\sqrt{s};a)} ,
    \label{eq:TChU}
\end{align}
where $V(\sqrt{s})$ is the kernel interaction constrained by chiral symmetry, 
which is a real function expressing dynamical contributions other than the 
$s$-channel unitarity. The function $G(\sqrt{s};a)$ is the once subtracted 
dispersion integral of the phase-space function $\rho(\sqrt{s})$, with which the 
unitarity of the amplitude is maintained through the optical theorem: 
$\im T^{-1}(\sqrt{s};a)=-\im G(\sqrt{s};a)=\rho(\sqrt{s})/2$. While the imaginary
part of $G(\sqrt{s};a)$ is given by the phase space, the real part depends on the
subtraction constant $a$.

We may identify the dispersion integral $G(\sqrt{s};a)$ as the loop function with
dimensional regularization. In this case, Eq.~\eqref{eq:TChU} is considered to be
the solution of the algebraic Bethe-Salpeter equation and the subtraction 
constant $a$ plays a similar role with the cutoff parameter in the loop function.
Through the order by order matching, the interaction kernel $V(\sqrt{s})$ is 
determined by chiral perturbation theory~\cite{Oller:2000fj}. At leading order, 
$V(\sqrt{s})$ is given by the $s$-wave interaction of the Weinberg-Tomozawa (WT) 
term
\begin{align}
    V(\sqrt{s})
    =&V_{\text{WT}}(\sqrt{s})
    =-\frac{C}{2f^2}
    (\sqrt{s}-M_T)
    \label{eq:WTterm} ,
\end{align}
where $C$, $M_T$ and $f$ are the group theoretical factor, the baryon mass, and 
the meson decay constant, respectively. With the leading order WT term, this 
framework is almost equivalent to the old coupled-channel works with vector meson
exchange potential~\cite{Dalitz:1967fp}. Based on chiral perturbation theory, it 
is now possible to include higher order correction 
systematically~\cite{Kaiser:1995eg,Lutz:2001yb}.

In the framework of N/D method, the CDD pole contribution should be included in 
the kernel interaction $V(\sqrt{s})$, except for the poles at infinity. Indeed, 
it is known that the contribution from the genuine states can be introduced in 
the interaction kernel $V(\sqrt{s})$, \textit{via} an explicit resonance 
propagator~\cite{Meissner:1999vr} or the contracted resonance contribution from 
the higher order Lagrangian~\cite{Oller:1998hw}. In the following, we demonstrate
that the CDD pole contribution can be embedded also in the loop function 
$G(\sqrt{s})$, and propose a method to extract such a hidden contribution by 
using a different renormalization scheme.

\section{CDD pole contribution in the loop function}
\label{sec:loop}

In the standard phenomenological studies, the interaction kernel $V(\sqrt{s})$ is
determined first, and then the subtraction constant $a$ has been fitted to 
reproduce experimental data. To illustrate the role of the subtraction constant 
in this approach, we argue the phenomenological amplitude in a schematic manner.
Suppose that we have the scattering amplitude $T_{\text{exp}}$ with enough 
experimental data. We may try to calculate this amplitude in chiral approach with
the leading order kernel $V^{(1)}$ and with the next-to-leading order kernel 
included $V^{(1)}+V^{(2)}$:
\begin{align}
    T^{(1)}(a^{(1)})
    =& \frac{1}{[V^{(1)}]^{-1}-G(a^{(1)})},
    \label{eq:amp1st} \\
    T^{(2)}(a^{(2)})
    =& 
    \frac{1}{[V^{(1)}+V^{(2)}]^{-1}-G(a^{(2)})}.
    \label{eq:amp2nd}
\end{align}
The subtraction constant should be chosen independently in each scheme for a good
description of the resulting scattering amplitude. If we achieve the complete 
description $T^{(1)}=T^{(2)}=T_{\text{exp}}$, the amplitudes in two schemes 
should become equivalent. In Eq.~\eqref{eq:amp1st}, the effect of the different 
interaction kernels must be compensated by the difference of the subtraction 
constant.

The leading order interaction $V^{(1)}$ is the WT term in Eq.~\eqref{eq:WTterm}, 
so it is clear that this term has no $s$-channel resonance contribution. On the 
other hand, the higher order term $V^{(2)}$ can have the CDD pole contribution 
from the contracted resonance propagator, as is known by the studies of chiral 
perturbation theory. If this is the case, the subtraction constant $a^{(1)}$ 
effectively contains the CDD pole contribution in the loop function for the 
scheme of Eq.~\eqref{eq:amp1st}.

The above exercise implies the existence of the CDD pole contribution in the loop
function. As far as phenomenological aspects of the model are concerned, the 
existence of the CDD pole contribution in the loop function is not a problem at 
all. However, in order to move one step forward to study the origin of resonances
in this approach, we should follow a different strategy to make the CDD pole 
contribution in the model under control.

\section{Natural renormalization scheme}
\label{sec:natural}

For this purpose, we propose the ``natural renormalization scheme,'' in which the
hidden CDD pole contribution in the loop function $G(\sqrt{s};a)$ is visualized 
in the interaction kernel $V(\sqrt{s})$. This situation can be achieved by 
requiring
\begin{itemize}
    \item  no state exists below the meson-baryon threshold, and

    \item  the amplitude $T$ matches with the interaction kernel $V$ 
    at certain low energy scale.
\end{itemize}
The latter condition is based on the validity of chiral expansion for low energy 
kinematics. These conditions uniquely determine the subtraction constant in the 
natural renormalization scheme $a_{\text{natural}}$ such that the loop function 
should satisfy~\cite{Hyodo:2008xr}
\begin{align}
    G(\sqrt{s};a_{\text{natural}})=& 0 \quad \text{at} \quad \sqrt{s} = M_T .
    \label{eq:naturalG}
\end{align}
The condition~\eqref{eq:naturalG} was already proposed in a different context; in
Ref.~\cite{Lutz:2001yb} the matching with the $u$-channel scattering amplitude 
was emphasized, and the matching with chiral low energy amplitude was argued in 
Refs.~\cite{Igi:1998gn,Meissner:1999vr}. Our point is to regard this condition as
the exclusion of the CDD pole in the loop function, based on the consistency with
the negativeness of the loop function. 

\section{Energy scale of the natural renormalization scheme}
\label{sec:scale}

Let us discuss the typical energy scale for the condition~\eqref{eq:naturalG}. In
Ref.~\cite{Oller:2000fj}, a ``natural'' value for the subtraction constant was 
estimated to be $a(\mu)\sim -2$, through the comparison of the loop function of 
dimensional regularization with that of three-momentum cutoff $q_{\text{max}}$ 
under nonrelativistic expansion, putting $q_{\text{max}}=\mu= 630$ MeV and the 
average target baryon mass $M_T= 1.15$ GeV.

This is different from our value of $a_{\text{natural}}$, practically and 
conceptually. The ``natural'' value in Ref.~\cite{Oller:2000fj} is obtained by 
fixing the typical energy scale of the meson-baryon scattering to be 630 MeV. In 
the present context, the condition~\eqref{eq:naturalG} is obtained by excluding 
the CDD poles the loop function, without introducing an explicit energy scale 
(such as $\sim$ 630 MeV). Therefore, it is not guaranteed that we obtain the 
``natural size'' $a(\mu)\sim -2$ in our natural renormalization scheme.

\begin{figure*}[btp]
    \centering
    \includegraphics[width=8cm,clip]{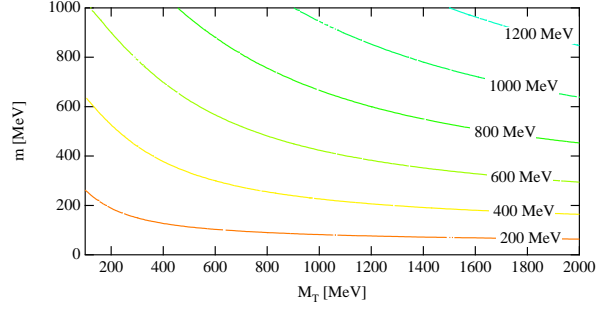}
    \label{fig:scale}
    \caption{
    Contour plot of the typical energy scale $q_{\text{max}}$ corresponding to 
    the natural subtraction constant $a_{\text{natural}}$ is shown as functions 
    of the masses $M_T$ and $m$.
    }
\end{figure*}

In this respect, it is instructive to study the typical energy scale given in
Eq.~\eqref{eq:naturalG}. As discussed in Ref.~\cite{Oller:2000fj}, it is not 
possible to match the real part of the loop functions in different regularization
in whole energy region, since they have different $\sqrt{s}$ dependence. Here we
estimate the scale of the loop function by matching it with three-momentum cutoff
regularization at threshold:
\begin{equation}
    G^{3d}(\sqrt{s}=M_T+m;q_{\text{max}})
    =G^{\dim}(\sqrt{s}=M_T+m,a_{\text{natural}}) .
    \label{eq:scale}
\end{equation}
Since $a_{\text{natural}}$ is given by Eq.~\eqref{eq:naturalG}, this equation 
determines the value of $q_{\text{max}}$ for given $m$ and $M_T$. The results are
shown in Fig.~\ref{fig:scale}. As seen in the figure, for $\bar{K}N$ scattering 
($m\sim 496$ MeV, $M_T\sim 939$ MeV), the typical scale is around $630$ MeV in 
accordance with the estimation in Ref.~\cite{Oller:2000fj}. On the other hand, 
when the pion is scattered as in $\pi\Sigma$ channel ($m\sim 138$ MeV, 
$M_T\sim 1193$ MeV), the scale of the natural renormalization is as small as 
$200$ MeV. One should keep in mind that the natural renormalization scheme may 
introduce an energy scale different from the scale of our interest.

\section{Interpretation of phenomenological model}
\label{sec:interpret}

Using the natural renormalization scheme, we can study the origin of the 
resonances in chiral dynamics. Let us assume that we have enough experimental 
data for the system of interest from the low-energy to the resonance-energy 
region. In the conventional phenomenological approaches, we choose the 
interaction kernel $V$ as the leading order WT term,
\begin{equation}
   T(\sqrt s; a_{\rm pheno}) = \frac{1}{V^{-1}_{\text{WT}}(\sqrt s) 
   - G(\sqrt s; a_{\rm pheno})}
   \label{eq:ampPheno} ,
\end{equation}
where the subtraction constant $a_{\text{pheno}}$ in the loop function $G$ is a 
free parameter to reproduce experimental data and takes care of the contributions
that are not included in the interaction kernel $V_{\text{WT}}$. We call this 
procedure the phenomenological renormalization scheme. 

On the other hand, the natural renormalization scheme fixes the subtraction 
constant such that in the resulting loop function there is no contribution from 
states below the threshold. To achieve the equivalent scattering amplitude, a 
different interaction kernel $V_{\rm natural}$ will be required:
\begin{equation}
   T(\sqrt s; a_{\rm natural}) = \frac{1}{V_{\rm natural}^{-1}(\sqrt s) - 
   G(\sqrt s; a_{\rm natural})}
   \label{eq:ampNatural} ,
\end{equation}
with the subtraction constant $a_{\rm natural}$. 

The scattering amplitude $T$ should equivalently be reproduced by both 
renormalization schemes. Thus, equating the denominators of 
Eqs.~\eqref{eq:ampPheno} and \eqref{eq:ampNatural}
\begin{eqnarray}
    V^{-1}_{\text{natural}}(\sqrt s) - G(\sqrt s; a_{\rm natural}) 
    =V^{-1}_{\text{WT}}(\sqrt s) - G(\sqrt s; a_{\rm pheno}) \ ,
\end{eqnarray}
we obtain the  effective interaction kernel $V_{\rm natural}$ in the natural 
renormalization scheme as
\begin{align}
    V_{\text{natural}}(\sqrt s) 
    = &-\frac{C}{2f^2}(\sqrt{s}-M_T)+\frac{C}{2f^2}
    \frac{(\sqrt{s}-M_T)^2}{\sqrt{s}-M_{\text{eff}}} ,
    \label{eq:pole} 
\end{align}
with an effective mass
\begin{align}
    M_{\text{eff}}\equiv &
    M_T-\frac{16\pi^2f^2}{CM_T\Delta a} ,\quad
    \Delta a = a_{\text{pheno}} - a_{\text{natural}}
    \label{eq:effectivemass} .
\end{align}
The expression~\eqref{eq:pole} indicates that the interaction kernel 
$V_{\text{natural}}(\sqrt s)$ can have a pole in the $s$-channel scattering 
region with an attractive interaction $C>0$ and a negative value for $\Delta a$. 
Note that the energy dependence of each term of Eq.~\eqref{eq:pole} is consistent
with the chiral expansion, since the pole term is quadratic in powers of the 
meson energy $\omega \sim \sqrt s - M_{T}$.

The relevance of the second term of Eq.~\eqref{eq:pole} depends on the 
scale of the effective mass $M_{\rm eff}$, which is obtained by the 
difference of the phenomenological and natural subtraction constants
$\Delta a$. If $\Delta a$ is small, the effective pole mass $M_{\text{eff}}$
becomes large. In this case, the second term of Eq.~\eqref{eq:pole} can be 
neglected or gives smooth energy dependence in the resonance energy region 
$\sqrt{s} \sim M_{T}+m \ll M_{\text{eff}}$. If the difference $\Delta a$ is
large, the effective mass $M_{\rm eff}$ gets closer to the threshold and the pole
contribution is no longer negligible. This means that the use of a negative 
$\Delta a$ with large absolute value is equivalent to the introduction of a pole 
in the chiral Lagrangian. We therefore consider that the pole in the effective 
interaction \eqref{eq:pole} is a source of the physical resonances in this case. 
In this way, we find a possible seed of the resonance in the loop function, even 
if we use the leading order chiral interaction.

In the phenomenological scheme~\eqref{eq:ampPheno}, the interaction kernel $V_{\text{WT}}$ does not 
include the CDD pole contribution, while in the natural
scheme~\eqref{eq:ampNatural} the loop function $G$ does not contain the CDD 
pole, as discussed in the previous section. Therefore, when the physical 
amplitude contains the CDD pole contribution, the effect is attributed to 
$G(\sqrt s; a_{\rm pheno})$ in the phenomenological scheme, while to 
$V_{\text{natural}}(\sqrt s) $ in the natural scheme. Indeed, we have 
demonstrated that $V_{\text{natural}}(\sqrt s)$ contains a resonance 
propagator. In the limit $\Delta a\to 0$, the two schemes agree with each 
other, which corresponds to the amplitude compatible with the meson-baryon 
picture of resonances, as explained in Sec.~\ref{sec:natural}. 

\section{Numerical analysis}\label{sec:numerical}

Let us apply the above method to physical baryon resonances in meson-baryon 
scattering. We consider meson-baryon scatterings in $S=-1$ and $I=0$ channel and 
$S=0$ and $I=1/2$ channel, where the $\Lambda(1405)$ and the $N(1535)$ resonances
are generated, respectively. For these channels, the phenomenological subtraction
constants $a_{\text{pheno},i}$ can be found in Refs.~\cite{Hyodo:2002pk,
Hyodo:2003qa} which are based on the results in Refs.~\cite{Oset:2001cn,
Inoue:2001ip}. In these models, the scattering observables such as cross sections
and phase shifts are well reproduced by the interaction kernel of the WT term. At
the same time, according to Eq.~\eqref{eq:naturalG}, we obtain the natural values
of the subtraction constants $a_{\text{natural},i}$ by setting $G(M_N)=0$ for all
channels.

\begin{figure*}[btp]
    \centering
    \includegraphics[width=16cm,clip]{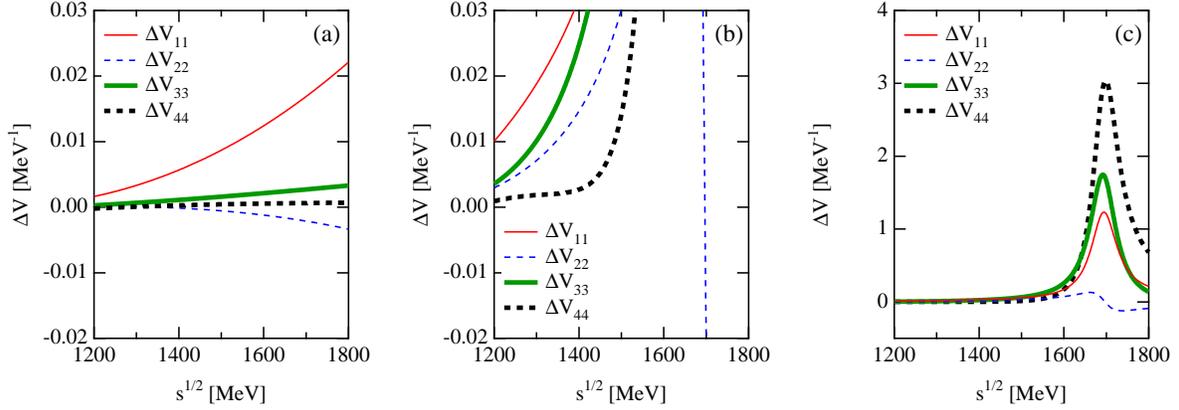}
    \label{fig:Eff}
    \caption{
    Deviations of the effective interactions from the Weinberg-Tomozawa term,
    (a) $S=-1$ channels,
    (b) enlargement of panel (c), 
    (c) $S=0$ channels. 
    The channels 1--4 correspond to 
    $\bar{K}N$, $\pi\Sigma$, $\eta \Lambda$, and $K\Xi$ for $S=-1$ channels,
    and to 
    $\pi N$, $\eta N$, $K \Lambda$, and $K\Sigma$ for $S=0$ channels, 
    respectively.
    }
\end{figure*}%

We evaluate the effective interaction in the natural renormalization scheme and 
extract the pole positions in the kernel. The nearest pole of the effective 
interaction in each channel is given by\footnote{With $n$-coupled channels, the 
effective interaction has $n$ poles, and a pair of complex poles can also 
appear~\cite{Hyodo:2008xr}.}
\begin{align}
    z_{\text{eff}}^{N^*} &= 1693 \pm 37 i \text{ MeV}  
    ,\quad
    z_{\text{eff}}^{\Lambda^*} 
    \sim 7.9 \text{ GeV} 
    \label{eq:poleeff} .
\end{align}
It is observed that the pole for the $N(1535)$ lies in the energy region of
resonance, while the pole for the $\Lambda(1405)$ is obviously out of the scale 
of the physics of the resonance. The influence of the pole~\eqref{eq:poleeff} can
be clearly seen in the diagonal components of the second term of 
Eq.~\eqref{eq:pole} (Fig.~\ref{fig:Eff}). We observe that the contributions are 
small for the $S=-1$ channels, whereas there is a bump structure at around 1700 
MeV in the $S=0$ channel, originate from the pole~\eqref{eq:poleeff} 
[Fig.~\ref{fig:Eff}(c)]. This result indicates that the $\Lambda(1405)$ is 
largely dominated by the component of the dynamical meson and baryon, while the 
$N(1535)$ may require some CDD pole contribution.

\begin{figure}[tbp]
    \centering
    \includegraphics[width=8cm,clip]{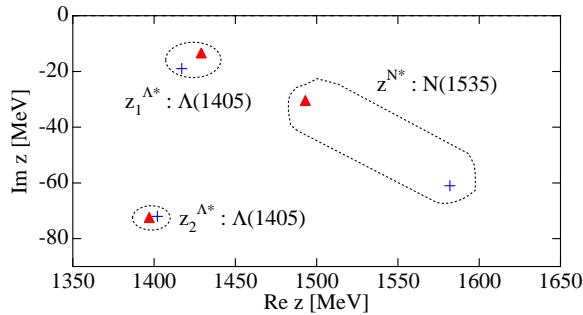}
    \caption{\label{fig:pole}
    Pole positions of the meson-baryon scattering amplitudes. The crosses denote 
    the pole positions in the natural renormalization with the WT interaction 
    (dynamical component); the triangles stand for the pole positions with the 
    phenomenological amplitude. $z_1^{\Lambda^*}$ and $z_2^{\Lambda^*}$ are the 
    poles for the $\Lambda(1405)$ in the $S=-1$ scattering amplitude, and 
    $z^{N^*}$ is the pole for the $N(1535)$ in the $S=0$ amplitude.}
\end{figure}

Next we consider the idealized purely dynamical components, which can be obtained
by adopting the WT term for the interaction kernel and the natural subtraction 
constant. In this case, there is no free parameter in the model and we can 
calculate the pole positions:
\begin{align}
    z^{N^*} &= 1582 - 61 i \text{ MeV}  ,
    \label{eq:naturalN}
    \\
    z_1^{\Lambda^*} &= 1417 - 19 i \text{ MeV}, \quad
    z_2^{\Lambda^*} = 1402 - 72 i \text{ MeV} .
    \label{eq:naturalL}
\end{align}
Note that in chiral dynamics the $\Lambda(1405)$ is described as the 
two poles in the complex energy plane~\cite{Jido:2003cb,Hyodo:2007jq}.
These can be compared with the pole positions in the phenomenological 
amplitude:
\begin{align}
    z^{N^*} &= 1493 - 31 i \text{ MeV} , 
    \label{eq:phenN} \\
    z_1^{\Lambda^*} &= 1429 - 14 i \text{ MeV}, \quad
    z_2^{\Lambda^*} = 1397 - 73 i \text{ MeV} ,
    \label{eq:phenL}
\end{align}
which corresponds to the physical resonances. We plot the pole positions in 
Fig.~\ref{fig:pole}. The poles for the $\Lambda(1405)$ appear in the close 
positions 
for the dynamical component~\eqref{eq:naturalL} and the physical 
one~\eqref{eq:phenL}. This again indicates the dominance of the meson-baryon 
component in the $\Lambda(1405)$. On the other hand, the pole for the $N(1535)$ 
moves to 
the higher energy when we use the natural values. Although the dynamical 
component generates a state by itself, the physical $N(1535)$ requires some more 
contributions, which is expressed as the pole in the effective 
interaction~\eqref{eq:poleeff} in the natural scheme. The comparison in 
Fig.~\ref{fig:pole} also indicates the dynamical nature of the $\Lambda(1405)$ 
and the sizable CDD pole contribution for the $N(1535)$.

\section{Conclusions}\label{sec:conclude}

In this report, we have discussed the origin of the resonances in chiral 
dynamics. From the viewpoint of the renormalization, we point out that the CDD 
pole contribution can be accommodated in the loop function, whose effect was not 
clear in the standard phenomenological fitting scheme. To avoid this kind of 
ambiguity of the interaction kernel, we construct the ``natural renormalization''
scheme for the loop function in which the CDD pole contribution is excluded. We 
show that it is possible to visualize the CDD pole contribution in the 
interaction kernel, from which the information of the origin of the resonances 
can be clearly extracted.

We analyze the $S=-1$ and $S=0$ meson-baryon scatterings in which the 
$\Lambda(1405)$ and the $N(1535)$ are dynamically generated. We find that 
the physical $\Lambda(1405)$ can be well reproduced when the leading order WT 
interaction is used as for the kernel of the scattering equation, while the 
$N(1535)$ requires a substantial correction in addition to the WT term, 
especially a pole singularity at around 1700 MeV. These facts indicate 
that the $\Lambda(1405)$ can be mainly described by a dynamical state of the 
meson-baryon scattering, which is consistent with the analysis of the $N_c$ 
scaling~\cite{Hyodo:2007np,Roca:2008kr} and the estimation of the electromagnetic
size~\cite{Sekihara:2008qk}. On the other hand, the $N(1535)$ may have an 
appreciable component beyond the present model space of meson-baryon two-body 
coupled channels. This could be, for instance, conventional three-quark state, 
correlated five-quark state, chiral partner of the ground state nucleon, or 
dynamical vector-meson plus baryon channels. Further investigation is called for 
the clarification of the structure of the $N(1535)$ resonance.

\section*{Acknowledgement}

This work was partly supported by the Global Center of Excellence Program by 
MEXT, Japan through the Nanoscience and Quantum Physics Project of the Tokyo 
Institute of Technology, and by the Grant-in-Aid for Scientific Research from 
MEXT (No.20028004, No.19540297). This work was done under the Yukawa 
International Program for Quark-hadron Sciences (YIPQS).

\end{document}